\documentclass[11pt]{article}
\usepackage[]{graphicx,float,latexsym,times}
\usepackage{amsfonts,amstext,amsmath,amssymb,amsthm}
\usepackage{amsmath,mathrsfs}
\usepackage{caption2}

\long\def\symbolfootnote[#1]#2{\begingroup%
\def\thefootnote{\fnsymbol{footnote}}\footnote[#1]{#2}\endgroup}

\usepackage{titlesec} 

\titleformat{\section}{\large\bfseries}{\thesection.}{.5em}{}
\titlespacing*{\section}{0pt}{*3}{*2}
\titleformat{\subsection}{\normalfont\bfseries}{\thesubsection.}{.5em}{}
\titlespacing*{\subsection} {0pt}{*3}{*2}
\titleformat{\subsubsection}{\normalfont\bfseries}{\thesubsubsection.}{.5em}{}
\titlespacing*{\subsubsection} {0pt}{*3}{*2}

\theoremstyle{plain} 
\newtheorem{theorem}{Theorem}[section]

\theoremstyle{definition} 

\newtheorem{remark}{Remark}[section]
\newtheorem{proposition}[theorem]{Proposition}

\numberwithin{equation}{section} 

 \usepackage[authoryear]{natbib}

\begin{document}


 \title{\textbf{\Large A class of sequential multi-hypothesis tests}}
 
 
  \maketitle

%
\author{
\begin{center}
\vskip -1cm

\textbf{\large Andrey Novikov}

 Metropolitan Autonomous University, 
Mexico City, Mexico

\end{center}
}

\symbolfootnote[0]{\normalsize Address correspondence to A. Novikov,
 Universidad Aut\'onoma Metropolitana, Unidad Iztapalapa, Avenida Ferrocarril  San Rafael Atlixco, 186, col. Leyes de Reforma 1A Secci\'on, C.P. 09310, Cd. de M\'exico, Mexico,
 E-mail: an@xanum.uam.mx}

{\small \noindent\textbf{Abstract:} 
In this paper, we deal with  sequential testing of multiple hypotheses. 
In the general scheme of construction of optimal tests based on the backward induction,
we propose a modification which provides a  simplified (generally speaking, suboptimal) version of the optimal test, for any particular criterion of optimization. We call this DBC version (the one with Dropped Backward Control) of the optimal test. 
In particular, 
for the case of two simple hypotheses,  dropping backward control in the Bayesian test produces  the classical  sequential probability ratio test (SPRT). Similarly, dropping backward control in the modified Kiefer-Weiss solutions  produces   Lorden's 2-SPRTs . 

In the case of more than two hypotheses, we obtain  in this way new classes of sequential multi-hypothesis tests, and  investigate their properties.
The efficiency of the DBC-tests is evaluated with respect to the optimal Bayesian multi-hypothesis test and with respect to the matrix sequential probability ratio test (MSPRT) by Armitage.
In a multihypothesis variant of the Kiefer-Weiss problem for binomial proportions 
the performance of the  DBC-test is numerically compared with that of the exact solution.
In a model of normal observations with a linear trend, 
the  performance of of the DBC-test is numerically compared with  that of the MSPRT. 
Some other numerical examples are presented.

In all the cases the proposed  tests exhibit a very high efficiency with respect to the optimal tests (more than 99.3\% when sampling from Bernoulli populations) and/or with respect to the MSPRT (even outperforming the latter in some scenarios).
}

\section*{ACRONYMS AND NOMENCLATURE}
\begin{tabular}{lp{9
cm}}
ESS& Expected sample size\\
SPRT& Secuential probability ratio test\\
MSPRT& Matrix sequential probability ratio test \citep{Armitage}\\
2-SPRT& Lorden´s 2-SPRT\\
$\psi=(\psi_1,\psi_2,\dots,\psi_n,\dots)$& Stopping rule\\
$I_{A}(x)$& Indicator function of event A\\
$k$& Number of hypotheses to be tested\\
$\phi_n^i$ &Conditional probability to accept hypothesis $H_i$ at stage $n$\\
$\phi_n=(\phi_n^1,\dots,\phi_n^k)$& Conditional probabilities to accept  $H_1,\dots, H_k$ at stage $n$\\
$\phi=(\phi_1,\phi_2,\dots,\phi_n,\dots)$& (Terminal) decision rule\\
$\langle \psi,\phi\rangle$&  Sequential test\\
$X_1,X_2,\dots, X_n,\dots$& Stochastic process of observations\\
$H_i:\theta=\theta_i$& $i$-th hypothesis about the underlying process\\
$\theta$ &Parameter of the distribution of the underlying process\\
$\tau_\psi$ & Stopping time generated by  stopping rule $\psi$\\
$s_n^\psi=(1-\psi_1)(1-\psi_2)\dots(1-\psi_{n-1})\psi_n$& Conditional probability to stop at stage $n$\\
$P_\theta$ & Distribution of the process when $\theta$ is the true value of parameter\\
$E_\theta(\cdot)$ & Expected value when $\theta$ is the true value of parameter\\
$\vartheta_i$& Parameter values to use for evaluation of the  expected sample size \\
$K$ & Number of  ESSs to be weighted in the minimization criterion\\
 $\alpha_{ij} (\psi,\phi)$& Error probability of test $\langle\psi,\phi\rangle$ (to accept $H_j$ given that $H_i$ is true)\\
 $\alpha_{i} (\psi,\phi)$& Error probability of test $\langle\psi,\phi\rangle$ (to reject $H_i$ given that $H_i$ is true)\\
 $C_{\gamma, \vartheta}(\tau_\psi)$ & Weighted ESS, to be minimized \\
$\gamma_i$& Weight for ESS calculated under $\vartheta_i$\\

 $\lambda_{ij}$& Lagrangian multipliers for $\alpha_{ij} (\psi,\phi)$\\
 $\lambda_{i}$& Lagrangian multipliers for $\alpha_{i} (\psi,\phi)$
\\
$L(\psi,\phi)$& Lagrangian function for test $\langle\psi,\phi\rangle$\\
$f_\theta^ n=f_\theta^ n(x_1,\dots,x_n)$ & Radon-Nikodym derivative of the distribution of $(X_1,X_2,\dots, X_n)$ with respect to a product-measure $\mu^n$ \\
$f^n_{\gamma\vartheta}=\sum_{i=1}^k\gamma_i f_{\vartheta_i}^n$ &Weighted distribution to be used for ESS minimization

\end{tabular}

\section{Introduction}

In this paper, we deal with sequential multi-hypothesis testing.
The Bayesian setting is  most commonly used for construction of optimal multi-hypothesis tests \citep[see, for example,][among others]{blackwell, Baum, Tartakovsky2014}. 
The distinctive feature of the Bayesian approach is the assumption that any one of the hypotheses comes up with some probability called {\em a priori}. In this way, the Bayesian risk is defined as a weighted sum of losses from  incorrect decisions plus the expected sample size, with the a priori probabilities as weights. The optimal test (called Bayesian) in this approach is that minimizing the Bayesian risk. The existence of optimal Bayesian tests is usually demonstrated applying general techniques of {\em dynamic programming} or {\em optimal stopping}  \citep[see, for example][for the case of two hypotheses]{Chow,Shiryaev}, and for the numerical approximations the {\em backward induction} is applied. 

The characteristics typically involved in the Bayesian risk are  error probabilities and expected sample size caculated under the hypotheses. 
At the same time, there is a number of statistical testing problems where the expected sample size should be  evaluated at parameter points distinct from the hypothesized ones. Probably the most notable is the so called modified Kiefer-Weiss problem, where the goal is to minimize the expected sample size under a parameter value located strictly between the hypothesized values \citep[see, for example,][]{Kiefer, Lai1973,Lorden76}.  Other examples can be found in \cite{eales}.
Some authors consider this type of problems Bayesian, too  \citep[for example][among others]{Weiss,eales}, but others don't \citep[see, for example][]{Lorden76,schmitz}. Whatever be the term, all these problems have much in common, because they incorporate the important characteristics of the tests
(like error probabilities, expected sample size, etc.) in a sole expression representing the   ``risk'' (or ``average loss'') of the whole statistical procedure. Unless their practical context is of  a  true Bayesian nature (with a real {\em a priori} distribution behind),  the use of Bayesian methodology in this type of problems is rather ``opportunistic''  \citep[as in the classical work of][for example]{waldwolfowitz}, because  the statisticians are in fact interested in the really  meaningful characteristics like error probabilities, etc.
This is why the Bayesian methods in these applications are often ``indirect'', because they require  re-calculation of the output characteristics in terms of the Bayesian input ``loss'' structure.

It should be noted nevertheless that there exist numerous hypothesis testing settings in the literature, which depart ``directly'' from the meaningful characteristics like error probability, statistical power, etc. \citep[see, for example][where many  references to previous work can be found]{Jennison}. Other  problems for more specific and/or more elaborated  models can  also be found in the literature  \citep[as in][for example]{chen2000, Silva2020}.

In this paper, we deal with the ``Bayesian'' model described above. In \cite{novikovmult}, a computer-oriented approach to constuction of optimal Bayesian tests was proposed. For the model of binary responses we developed a complete set of agorithms facilitating the evaluation of optimal tests and their characteristics for this problem, implemented them in the R  programming language and made them publicly available in the form of a GitHub repository. With the program at hand it is easy to find (numerically) the input values in the Bayesian problem in order to satisfy  the requirements on output quality.
This approach can be, in principle,  adapted to other exponential families of distribution, but requires some programming work to be done for any specitic family.

In this paper, we propose a modification of the optimal
 sequential tests which is quite simple because only affects the stopping rule in the way the ``backward induction'' part of the governing equations is omitted, while this is the most laborious part of the dynamic programming techniques. At the same time this modification is rather universal because it is applicable to virtually any stochastic model, at least to those with calculatable expressions for joint distributions. In addition, it is applicable on finite and infinite horizons and it requires only forward run for its evaluation. We call this modification ``optimal test with dropped backward control (DBC)''. We numerically  study the properties of the proposed tests, evaluating their characteristics and comparing them with those of the optimal tests. The problems  and models examined are: Bayesian testing between three simple hypotheses about binomial proportions,
 testing between two two-sided hypotheses about binomial proportions, Bayesian test for  three hypotheses about the mean of a normal distribution with a linear trend in the mean, and   the   Kiefer-Weiss testing between three hypotheses about binomial proportions. 


In the particular case of two hypotheses, this method provides the well-known class of sequential probability ratio tests (SPRTs) in the Bayesian setting, and  the class of 2-SPRTs by \cite{Lorden76} for the modified Kiefer-Weiss problem, as well as some other tests known from the literature. 
  
 For three or more hypotheses the proposed method generates a new class of sequential multi-hypothesis tests, whose properties we investigate in this paper.  

 The remainder if the paper is organized as follows.  Section \ref{Lagr}  explains the use of the method of Lagrange multipliers for optimal sequential multi-hypothesis testing. 
 Section 3 introduces the new class of multihypothesis tests called DBC (Dropped Backward Control) variants of the optimal tests and some of  their  properties are highlighted.
 In Section 4, five numerical examples of applications of the new tests are presented. 

\section{Optimal multi-hypothesis tests}\label{Lagr}

In this section, we describe a general method of construction of the optimal sequential multi-hypotheses tests. Essentially, this is a brief recount  of the full construction found in \cite{novikovMultiple} adapted to the present context.

Let $X_1,X_2,\dots,X_n,\dots$ be a sequence of random variables the statistician is receiving for analysis on the one-by-one basis. The distribution of the sequence is uniquely defined by a parameter $\theta$ the statistician wants to decide about, namely, which one of the $k$ hypotheses is true: $H_1$: $\theta=\theta_1$, $H_2$: $\theta=\theta_2$, $\dots$, or $H_k$: $\theta=\theta_k$, $k\geq 2$.

Let  $\langle\psi,\phi\rangle$ be a sequential multi-hypothesis test, $\tau_\psi$ its correponding stopping time,
$E_{\theta}\tau_\psi$ the expected sample size when $\theta$ is the true value of the parameter, $ \alpha_{ij}(\psi,\phi)$ (or $\alpha_{i}(\psi,\phi))$) the error probabilities  of the test  $\langle\psi,\phi\rangle$.  We refer to \cite{novikovMultiple} for the formal definitions.

The general purpose of the construction  is the minimization of the  expected sample size 
weighted over some set of parameter points.
Let $\vartheta_i$, be some parameter points, $i=1, \dots, K$,  $K\geq 1$, and $\gamma_i>0$ be some weights  (i. e. $\sum_{i=1}^K\gamma_i$=1).
Then we are interested  in minimizing 

\begin{equation}\label{9-1}
C_{\gamma, \vartheta}(\tau_\psi)=\sum_{i=1}^K\gamma_iE_{\vartheta_i}\tau_\psi
\end{equation}
under the conditions
\begin{equation}\label{9-1bis}
\alpha_{ij}(\psi,\phi)\leq \alpha _{ij},\;1\leq i\not=j\leq k\quad\text{(or}\; \alpha_i(\psi,\phi)\leq \alpha _{i}, \;1\leq i\leq k \text{)},
\end{equation}
where $\alpha _{ij}$ and $\alpha _{i}$ are some non-negative numbers.

The problem is solved applying the method of Lagrange multipliers. 

Let $\lambda_{ij}$ be some non-negative constants, $1\leq i\not=j\leq k$.
Then the problem of minimization of \eqref{9-1} under conditions on $\alpha_{ij}(\psi,\phi)$ reduces to that of minimization of
\begin{equation}\label{1-1}
 L(\psi,\phi)=C_{\gamma ,\vartheta}(\tau_\psi)+\sum_{1\leq i\not=j\leq k}\lambda_{ij}\alpha_{ij}(\psi,\phi).
\end{equation}
 Quite analogously, the problem of minimization under restrictions on $\alpha_i(\psi,\phi)$ reduces to minimization of
\begin{equation}\label{2-1}
 L(\psi,\phi)=C_{\gamma, \vartheta}(\tau_\psi)+\sum_{1\leq i\leq k}\lambda_{i}\alpha_{i}(\psi,\phi)
\end{equation}
with some Lagrangian multipliers $\lambda_i\geq 0$, $1\leq i\leq k$ \citep[see][]{novikovMultiple}. It is easy to see that \eqref{2-1} is a particular case of \eqref{1-1} when $\lambda_{ij}=\lambda_i$ for all $j\not=i$, so we focus on the problem of minimization of \eqref{1-1}.

When $\theta_i=\vartheta_i$ for $i=1,2,\dots , k=K$ the Lagrangian function \eqref{1-1} can also be interpreted as a Bayesian risk corresponding to the a priori distribution $\{\gamma_1,\gamma_2,\dots,\gamma_k\}$ on $\{\theta_1, \dots, \theta_k\}$. 

To characterize the tests minimizing  \eqref{1-1}, an additional assumption is needed. Let us assume that 
for each $n$ and each $\theta$ the vector $(X_1,\dots,X_n)$ has a density $f_\theta^ n=f_\theta^ n(x_1,\dots,x_n)$ (Radon-Nikodym derivative) with respect to a product-measure $\mu^n$ ($n$ times $\mu$ by itself, where $\mu $ is a $\sigma$-finite measure). Let us also denote $f^n_{\gamma\vartheta}=\sum_{i=1}^k\gamma_i f_{\vartheta_i}^n$.

Let now 
\begin{equation}
 v_n=v_n(x_1,\dots,x_n)=\min_{1\leq j\leq k}\sum_{i:i\not=j}\lambda_{ij} f_{\theta_i}^n(x_1,\dots,x_n)
\end{equation}
Then the optimal decision rule $\phi$ can be defined as
\begin{equation}\label{3-2}
 \phi_n^j=0\;\text{whenever}\;\sum_{i:i\not=j}\lambda_{ij} f_{\theta_i}^n>v_n,
\end{equation}
so that, with this $\phi$,
\begin{equation}\label{1-2}
L(\psi,\phi)\geq L(\psi)=\sum_{n=1}^\infty \int s_n^\psi(nf_{\gamma\vartheta}^n+v_n)d\mu^n
\end{equation}
and  the problem is reduced to that of minimization of the right-hand side of
\eqref{1-2} over all stopping rules $\psi$ (see \cite{novikovMultiple}).

This latter problem is first solved over the class of truncated stopping rules.
Let $\mathcal S^N$ be the class of stopping rules such that 
$(1-\psi_1)(1-\psi_2)\dots(1-\psi_{N})=0$.
Then the the optimal stopping rule in $\mathcal S^N$ is constructed as follows.

Starting from $V_N^N\equiv v_N$, define recursively over $n=N,\dots,2$
\begin{equation}\label{backward}
 V_{n-1}^N=\min\{v_{n-1},f_{\gamma\vartheta}^{n-1}+\mathcal I_n V_n^N \},
\end{equation}
where   $\mathcal I_n $ is an operator defined for any non-negative measurable function $ v=v(x_1,\dots,x_n)$ as
$$
\mathcal I_n v=(\mathcal I_nv)(x_1,\dots,x_{n-1})=\int v(x_1,\dots,x_n)d\mu(x_n).
$$
Equation \eqref{backward} is what usually called ``backward induction'' equation.

Then for any $\psi\in\mathcal S^N$
\begin{equation}\label {1-3}
 L(\psi)\geq 1+\mathcal I_1 V_1^N,
\end{equation}
and there is an equality in \eqref{1-3} when
\begin{equation}\label{1-4}
 \psi_n=I_{\{v_n\leq f_{\gamma\vartheta}^n+\mathcal I_{n+1}V_{n+1}^N\}}
\end{equation}
for all $n=1,2,\dots,N-1$. Thus, stopping rule \eqref{1-4} solves the problem of minimization of $L(\psi)$ in the class $\mathcal S^N$ of truncated stopping rules. The stopping rule can be arbitrarily randomized between the samples for which there is an equality in the inequality under the indicator function in \eqref{1-4} \citep{novikovMultiple}.

Under very mild condition \citep[see, for example,][Remark 7]{novikovMultiple}, the solution of the problem of minimization of
\eqref{1-2} is obtained by letting $N\to\infty$.

For any $n\geq 1$ and $N\geq 1$ $V_n^{N+1}\leq V_n^N$, so there exists $\lim_{N\to\infty} V_n^N=V_n$, so it follows from \eqref{1-3} that
\begin{equation}\label {1-3bis}
 L(\psi)\geq 1+\mathcal I_1 V_1,
\end{equation}
 for all stopping rules $\psi$, and 
there is an equality in \eqref{1-3bis} if 
\begin{equation}\label{1-5}
 \psi_n=I_{\{v_n\leq f_{\gamma\vartheta}^n+\mathcal I_{n+1}V_{n+1}\}}
\end{equation}
for all $n=1,2,\dots$. Thus, \eqref{1-5} defines an optimal stopping rule. Again, it can be arbitrarily randomized in case there is an equality in the inequality in \eqref{1-5}. The details of this result can be found in \cite{novikovMultiple}.

As a concluding remark, let us note that the problem of minimization of \eqref{1-1} has a trivial decision in case
$$
\sum_{i:i\not= j}\lambda_{ij}\leq 1 \quad \text{for some} \quad j=1,2 \dots k.
$$
Indeed, in this case the trivial test $\langle\psi_0,\phi_j\rangle$ that always (without any observation) takes the decision to accept $H_j$, has $\alpha_{ij}(\psi_0,\phi_j)=1$ for all  $i\not=j$ and $\alpha_{ji}(\psi_0,\phi_j)=0$ for all $i\not=j$, so
$$
\sum_{i,j:i\not=j}\lambda_{ij}\alpha_{ij}(\psi_0,\phi_j)=\sum_{i:i\not= j}\lambda_{ij}\leq 1+\mathcal I_1V_1
$$
which means the trivial test performs better than the best test  among those taking at least one observation.  

Therefore, when it comes to minimization of $L(\psi,\phi)$ (or Bayesian risk) we will always assume that
\begin{equation}
\sum_{i:i\not=j}\lambda_{ij}>1 \quad\text{for all}\quad j=1,\dots,k,
\end{equation}
In particular, in the case of  $k=2 $ hypotheses, $\lambda_{12}(=\lambda_1)>1$ and $\lambda_{21}(=\lambda_2)>1$.
\section{Proposed modification}

The idea of the modification of the optimal multi-hypothesis test we propose in this paper is suggested  by the form of optimal stopping rule  \eqref{1-5}. We see from \eqref{1-5} that the optimal test stops (that is, $\psi_n=1$) when 
\begin{equation}\label{1-6}
 v_n\leq f_{\gamma\vartheta}^n
\end{equation}
(because  all $V_n$ are non-negative). This corresponds to ``dropping'' the term $\mathcal I_{n+1}V_{n+1}$ 
in \eqref{1-5} which is responsible, at step $n$, for
the optimum stopping due to the future behaviour  of the controlled process. 
Without any additional information about the exact form of $V_{n+1}$, a natural ``simplified'' way of  acting is to continue whenever \eqref{1-6} does not hold. This takes us to the stopping time $\tau_\psi =\min\{n\geq 1:v_n\leq f_{\gamma\vartheta}^n\}$ (see \eqref{1-6}).
$\tau_{\psi}$ stops at a later time than   the optimal one does, but has the advantage that it does not require any specification of the functions $V_n$, which   is normally  not an easy task.

 Taking into account the above arguments, we propose the following multi-hypothesis test for the general case of $k\geq 2$ hypotheses.

 Let for any natural  $n$
 \begin{equation}\label{1-7}
 \psi_n=1 \;\mbox{whenever}\;\min_{j=1,\dots,k}\sum_{i:i\not=j}\lambda_{ij}f_{\theta_i}^n\leq \sum_{i=1}^K\gamma_i f_{\vartheta_i}^n,
\end{equation} 
and
\begin{equation}\label{3-1}
 \phi_n^j=0 \;\mbox{whenever}\;\ \sum_{i:i\not=j}\lambda_{ij} f_{\theta_i}^n>\min_{l=1,\dots,k}\sum_{i:i\not=l}\lambda_{il} f_{\theta_i}^n
\end{equation}

The minimum on the right hand side of \eqref{3-1} is achieved for at least one $j$. If there is  more than one $j$ with this property,  the final decision may be obtained by an arbitrary randomization between  all of them. A randomization can also be applied in case there is an equality in \eqref{1-7}.

We will call the test  $\langle \psi,\phi\rangle$ defined by \eqref{1-7} -- \eqref{3-1} DBC-version of the optimal one,  because it is obtained by omitting the details of  the functions $V_n$ responsible for the exact form of the optimal control.

\begin{remark}{ It is  worth noting that the test defined by \eqref{1-7} -- \eqref{3-1} is applicable to statistical models where the observations    are  not necessarily independent or identically distributed. Restriction of \eqref{1-7} -- \eqref{3-1} to steps $1\leq n\leq N$ is also a DBC version of the optimal truncated 
 test \eqref{1-4} -- \eqref{3-2}.
 }
 \end{remark}

\subsection{Some properties of DBC-test for two hypotheses}\label{2hyp}
Let us show that, in  some particular cases of two hypotheses, the DBC-test defined in \eqref{1-7} -- \eqref{3-1} reduces to  very well known sequential tests.

 Let us first look at the classical case of two simple hypotheses in the Bayesian setting (i. e., $\vartheta_i=\theta_i$, $i=1,2$).
 It is easy to see that in this case stopping according to \eqref{1-6}  is equivalent to stopping when 
 $$\text{either}\quad\lambda_2 f_{\theta_2}^n\leq \gamma_1 f_{\theta_1}^n+\gamma_2f_{\theta_2}^n \quad\text{or}\quad 
 \lambda_1 f_{\theta_1}^n\leq \gamma_1 f_{\theta_1}^n+\gamma_2f_{\theta_2}^n,$$
 which is equivalent to
 $$
 \frac{ f_{\theta_2}^n}{ f_{\theta_1}^n}\leq \frac{\gamma_1}{\lambda_2-\gamma_2}=A<1
 \quad\text{and}\quad
 \frac{ f_{\theta_2}^n}{ f_{\theta_1}^n}\geq \frac{\lambda_1-\gamma_1}{\gamma_2}=B>1,
 $$
respectively. That is, our ``simplified'' procedure yields in fact  the classical stopping time   of an  SPRT 
$$
 \min\{n\geq 1:\frac{ f_{\theta_2}^n}{ f_{\theta_1}^n}\not\in(A,B)\}
$$
 introduced by A. Wald  at the very beginning of the sequential analysis \citep[see, for example,][]{Wald45}. 

 The decision rule  \eqref{3-2} in this case is to accept hypothesis $H_2$ when
 \begin{equation}\label{3-4}
  \frac{f_{\theta_2}^n}{f_{\theta_1}^n}\geq \frac{\lambda_1}{\lambda_2} 
 \end{equation}
and accept $H_1$ otherwise.

It is easy to see that
\begin{equation}\label{3-3}
A=\frac{\gamma_1}{\lambda_2-\gamma_2}< \frac{\lambda_1}{\lambda_2}< B=\frac{\lambda_1-\gamma_1}{\gamma_2}
\end{equation}
which implies that our  rule \eqref{3-4} takes the same decision  the SPRT does, when this latter stops. 

To prove \eqref{3-3} let us suppose that, by
the contrary, 
$$
\frac{\lambda_1}{\lambda_2}\leq\frac{\gamma_1}{\lambda_2-\gamma_2}, 
$$
say. Then 
$$
\lambda_1\lambda_2\leq\gamma_1\lambda_2+\gamma_2\lambda_1\leq \max\{\lambda_1,\lambda_2\}
$$
which contradicts to the fact that both $\lambda_1> 1$ and  $\lambda_2> 1$.
Analogously, the second inequality in \eqref{3-3} can be proved.

Thus, any Bayesian (as well as any SPRT) for two simple hypotheses is a particular case of the DBC-test \eqref{1-7} -- \eqref{3-1}.

Let us look now at the  Kiefer-Weiss problem. The problem is originally formulated by \cite{Kiefer} for two hypotheses and consists in minimizing $\sup_\theta E_\theta \tau_\psi$ over all tests subject to restrictions on the error probabilities. They suggested that the problem, in some cases,  can be reduced to minimization of $E_\vartheta\tau_\psi$ for a specific choice of $\vartheta$, under the same restrictions (this problem is called nowadays the modified Kiefer-Weiss problem).
This exactly matches the problem of minimization of $C_{\gamma, \vartheta}(\tau_\psi)$ with $K=1$ and $\gamma_1=1$ and $\vartheta_1=\vartheta$ (see \eqref{9-1}) under restrictions on $\alpha_i(\psi,\phi)$, $i=1,2$. Thus, our DBC-test \eqref{1-7} - \eqref{3-1} is applicable. 

Let us examine the particular form the DBC-test \eqref{1-7} - \eqref{3-1} acquires in this case.

The stopping time \eqref{1-7} provides
\begin{equation}\label{9-3}
 \tau_\psi=\min\{n\geq 1:\min\{\lambda_{1}f_{\theta_1}^n,\lambda_{2}f_{\theta_2}^n\}\leq  f_{\vartheta}^n\},
\end{equation}
and the decision is in favour of $H_1$ (i. e., $\phi_n=1$), whenever
\begin{equation}\label{9-4}
 \lambda_{1} f_{\theta_1}^n\geq \lambda_{2} f_{\theta_2}^n.
\end{equation}

It is easily seen that this is a variant of the 2-SPRT proposed by \cite{Lorden76}, where $A=1/\lambda_1$ and $B=1/\lambda_2$. The numerical results of \cite{Lorden76} show a very high efficiency of the 2-SPRT in the symmetric normal case. 

In  Section \ref{comp} (see Example V) we also suggest using the test \eqref{1-7} - \eqref{3-1}
in the multi-hypothesis version of the Kiefer-Weiss problem.

\begin{remark}{Taking into account the Bayesian  origin of the proposed test, let us represent it in the traditional Bayesian terms like the prior and posterior probabilities of hypotheses. This setting corresponds to the particular case of our proposed test \eqref{1-7} - \eqref{3-1}, when $\vartheta_i=\theta_i$, $i=1,2,\dots, k$, and $K=k$.

Interpreting $\gamma_i$, $i=1,\dots, k$ as the prior probabilities of the respective hypotheses $H_1$, $\dots$, $H_k$, let us denote $$\pi_i^n=\frac{\gamma_i f_{\theta_i}^n}{\sum_{j=1}^k \gamma_j f_{\theta_j}^n},\; i=1,\dots,k, $$
the posterior probabilities of the hypotheses, $n=1,2,\dots$, and let $\pi_i^0=\gamma_i$.


Then 
stopping rule \eqref{1-7}  gives 
\begin{equation}
 \tau_\psi=\min\{n\geq 1:\min_{j=1,\dots,k}\sum_{i:i\not=j}\lambda_{ij}\pi_i^n/\gamma_i\leq 1\},
\end{equation}
and the  condition of deciding in favour of $H_j$ in \eqref{3-1}  is equivalent to
\begin{equation}
 \sum_{i:i\not=j}\lambda_{ij} \pi_i^n/\gamma_i=\min_{l=1,\dots,k}\sum_{i:i\not=l}\lambda_{il} \pi_i^n/\gamma_i.
\end{equation}

 In this way, our proposed test \eqref{1-7} - \eqref{3-1} provides a simplified form (omitting the backward recursion) of  the Bayesian tests in the multi-hypothesis case. 
 }\end{remark}

\subsection{ A general stopping property}
We show in this section that, under rather general conditions,   $\tau_\psi<\infty$ with probability 1 under any hypothesis $H_j$, for $\psi$ defined in \eqref{1-7}. Let us denote $\Theta_1=\{\theta_i:1\leq i\leq k\}$, 
$\Theta_2=\{\vartheta_i:1\leq i\leq K\}$ (may coincide).

\begin{proposition}\label{p1}
Let us suppose that   $\vartheta\in \Theta_2$ is such that
${f_{\theta}^n}/{f_{\vartheta}^n}\to 0 $ in $P_{\vartheta}$- probability, as $n\to\infty$,
for all $\theta\not=\vartheta$ in $\Theta_1\cup\Theta_2$.
Then $P_{\vartheta}(\tau_\psi<\infty)=1$.
\end{proposition}

{\bf Proof.} Let us suppose that the conditions of Proposition \ref{p1} hold. 
Let $\vartheta=\vartheta_i\in \Theta_2$.
If $\vartheta_j\in \Theta_1$, let $l$ be such that $\theta_l=\vartheta_j$, otherwise let $l=1$. Then

\begin{equation}\label{5-1}
P_{\vartheta_j}(\tau_\psi>n)\leq P_{\vartheta_j}(\sum_{i:i\not=l}\lambda_{il}f_{\theta_i}^n> \sum_{i=1}^K\gamma_i f_{\vartheta_i}^n)=P_{\vartheta_j}(\sum_{i:i\not=l}\lambda_{il}\frac{f_{\theta_i}^n}{f_{\vartheta_j}^n}-\sum_{\vartheta_i\not=\vartheta_j}\gamma_i \frac{f_{\vartheta_i}^n}{f_{\vartheta_j}^n}>  \gamma_j).
\end{equation}
It follows from the condition of Proposition \ref{p1} that  the sums under the probability sign on the right-hand side of \eqref{5-1} tend to 0 in $P_{\vartheta_j}$- probability, which implies that $P_{\vartheta_j}(\tau_\psi>n)\to 0$ as $n\to\infty$, so $P_{\vartheta_j}(\tau_\psi=\infty)= 0$, that is, $P_{\vartheta_j}(\tau_\psi<\infty)=1$. $\Box$

In the case of independent and identically distributed (i.i.d.) observations
the conditions of Proposition \ref{p1} are satisfied when  the distributions  of $X$ for  any pair of distinct $\theta\in\Theta_1\cup\Theta_2$ are distinct. Indeed, in this case for any positive $\epsilon$
\begin{equation}\label{5-2}
P_{\vartheta}({f_{\theta}^n}/{f_{\vartheta}^n}>\epsilon)\leq E_{\vartheta} \left(\frac{f_{\theta}^n}{f_{\vartheta}^n}\right)^{1/2}\epsilon ^{-1/2}=
\left(\int (f_{\theta}f_{\vartheta})^{1/2}d\mu\right)^n\epsilon ^{-1/2}=(r_{\theta\vartheta})^n\epsilon ^{-1/2},
\end{equation}
where $r_{\theta\vartheta}=\int (f_{\theta}f_{\vartheta})^{1/2}d\mu<1$, given that $\mu(f_{\theta}\not=f_{\vartheta})>0$. Thus, the right-hand side of \eqref{5-2} tends to 0, as $n\to\infty$.
Moreover, it follows from   \eqref{5-2} that $\tau_\psi$ is exponentially bounded under these conditions, namely, that
$
P_{\vartheta}(\tau_\psi>n)\leq cr_\vartheta^n
$ with some $c$ for all $n\geq 1$, where  $r_\vartheta=\underset{\theta:\theta\not=\vartheta}{\text{max}} \;r_{\theta\vartheta}<1$.  Because of this,
$E_{\vartheta}\tau_\psi=\sum_{n=1}^\infty P_{\vartheta}(\tau_\psi\geq n)\leq 1+\sum_{n=2}^\infty cr_\vartheta^{n-1}=1+cr_\vartheta/(1-r_\vartheta)<\infty
$

For Markov-dependent observations, \cite{sueselbeck} showed that some SPRTs for testing  two simple hypotheses are not closed (in the sense their stopping times are infinite  with positive probability). Naturally, the conditions of Proposition \ref{p1} can not be satisfied in this case. On the other hand, there are non-trivial examples of  SPRTs which are closed, despite that the conditions of Proposition \ref{p1} do not hold, so the conditions of Proposition \ref{p1} are not necessary for being the test \eqref{1-7} --  \eqref{3-1} closed.

Even when no general conditions exist  which guarantee that stopping time \eqref{1-7} is finite (like those of Proposition \ref{p1}), there is still a way to use the DBC variant of the optimal test just truncating the stopping time  at some reasonable (and/or convenient, and/or required) level, i. e. using as the stopping time $\min\{N,\tau_\psi\}$ where $\psi$ is defined by \eqref{1-7} and $N$ is a finite horizon. The decision rule does not need to change in any way (i. e. \eqref{3-1} may be used).
 
 In this way, the proposed DBC-tests are quite universal: they can be used for testing any number of  hypotheses for independent or dependent observations, on the infinite or finite horizon. 

\subsection{Computational aspects}

The use of the method of Lagrange  in Section \ref{Lagr} requires the determination of  the  multipliers $\lambda_{ij}$ in such a way that the  test minimizing \eqref{1-1} complies with the restrictions \eqref{9-1}.  Because the (probably most)  essential element governing the optimal behaviour has been dropped, we can not expect the DBC version would (even approximately) be optimal, for these given $\lambda_{ij}$. Nevertheless, keeping in mind  that the rules of the DBC-test  still bear inside the   multipliers  $\lambda_{ij}$, we can try to use them as was initially intended: to vary  $\lambda_{ij}$  trying to make the error probabilities satisfy the restrictions. Just instead of the optimal test we now have the DCB-version to be manipulated in this way. There is no theoretical reason by which the resulting tests could have any attractive properties, but it may be encouraging that, in the case of two hypotheses, the resulting DCB-tests  provide highly efficient tests (see  Section \ref{2hyp}). More precisely,
making the modification to the optimal Bayesian  test, we obtain  an  SPRT, subject to the needed error probabilities, which is {\em optimal} under these conditions \citep[][]{waldwolfowitz}. And doing the same with the modified Kiefer-Weiss solution one obtains the Lorden's 2-SPRT, which has a very high  efficiency (exceeding 99\%) with respect to the {\em optimal} Kiefer-Weiss solution  \citep[see][]{Lorden76}. The examples we present in Section \ref{comp} provide more situations where DBC-tests are also highly  efficient.

As described above, to apply the proposed DBC-tests in the practice we need a routine for the numerical evaluation of  their characteristics (like error probabilities, expected sample size, etc.) for any set of parameters $\lambda_{ij},  1\leq i,j\leq k$, $i\not=j$, and $\gamma_i$, $\vartheta_i$, $1\leq i\leq k$.  
The algorithms developed in \cite{multihypothesisGit},  for the particular case of binomial model, enable all the needed computations both for the optimal and for the DBC version of multihypothesis tests. For  other statistical models, the numerical evaluation of the optimal tests requires  developing  algorithms implementing backward recursion (like those for two hypotheses in \cite{novikovfarkhshatovDiscrete, continuous}). In contrast,  the implementation of the proposed DBC-tests appears to be rather straightforward, at least while Monte Carlo simulations can be used for the  approximate evaluation of statistical characteristics of the tests. The DBC-tests themselves  normally will not cause computational complications in this sense, as  seen from    \eqref{1-7} -- \eqref{3-1}. 

For independent observations, \cite{liu} proposed computational algorithms for evaluating the characteristics of SPRTs  and multi-hypothesis  matrix sequential probability tests (MSPRT) 
 by \cite{Armitage}. It is likely that  the algorithms of \cite{liu} can be extended to  some classes of the DBC-tests.

\section{Computations and simulations results}\label{comp}

In this Section, we want to see numerical results related to applications of the proposed test \eqref{1-7} -- \eqref{3-1}. We use the program code in \cite{multihypothesisGit} for binary data models  in this section. The basic algorithms are described in \cite{novikovmult}, the function  {\tt DBCTest} for designing the DBC-test \eqref{1-7} -- \eqref{3-1} is newly added to the repository.
 \vspace{4pt}

{\flushleft\em Example I} \vspace{2pt}

First of all, let us revisit the example in \cite{novikovmult} to numerically compare the performance of the proposed test with that of the optimal (Bayesian) sequential multi-hypothesis  test. 

We refer to the case of testing three hypotheses $H_1:\theta=0.3$, $H_2:\theta=0.4$ and $H_3:\theta=0.5$ about the success probability $\theta$ of a Bernoulli distribution.  

In this case, all the evaluations of the performance characteristics can be made  using the algorithms  developed in \cite{novikovmult} and are carried out with double computer precision. 

For  a series of nominal error probabilities $\alpha_1=\alpha_2=\alpha_3=\alpha$, 
where $\alpha$ is any one of the numbers 0.1, 0.05, 0.025, 0.01, 0.005, 0.002, 0.001, 0.0005,  we found the optimal multi-hypothesis  tests which fit best (up to approx. 0.002 of the relative distance) the nominal value of $\alpha$. As a criterion of optimization we used the Bayesian criterion \eqref{2-1} with the uniform weights ($\gamma_1=\gamma_2=\gamma_3=1/3$). The optimization was conducted over $\lambda_i$,  $i=1,2,3$. The gradient-free general-purpose optimization method by \cite{neldermeadarticle} was used for finding the best fit, with respect to the relative distance
\begin{equation}
\max_i\{|\alpha_i(\psi,\phi)-\alpha_i|/\alpha_i\}.
\end{equation}

The same fitting procedure was then carried out for our  DBC-test, with the same vector of $\gamma$ weights, but over its own multipliers  $\lambda_{ij}=\lambda_i$ for $j\not=i$, $i=1,2,3$ (see \eqref{1-7} -- \eqref{3-1}). The trucation level of $N=3000$ was used for all the evaluations.

The fitted results are shown in Table 1. The efficiency is defined as the ratio between the minimum ESS calculated for Bayesian and for DBC-test.
In each column,   the values of the weighted ESS \eqref{9-1}  are presented (with $\vartheta_i=\theta_i$), both for the optimal test (first row) and for its DBC variant (second row). We observe a very high efficiency of the DBC-test (over 99.3\%) with a clear tendency of increasing for small values of $\alpha$. 

Yet another series of numerical experiments was carried out, in the same model but in a slightly more applied context. Let us imagine that a clinical trial of a cancer treatment is on the run where the proportion of ``regressions'' is observed, after the   treatment has been applied. The hypotheses are $H_1:\theta=0.1$ (low proportion of regressions), $H_2:\theta=0.3$ (moderate proportion of regressions) and $H_3:\theta=0.5$ (very high proportion). As usual in sequential clinical trials, there is an ethical issue in this case: if the treatment
is turning out to be non-efficient, it should be terminated as soon as possible. This may be controlled by assigning a higher weight to the expected sample size corresponding to the  no-efficiency hypothesis $H_3$: let $\gamma_3=0.8$ while $\gamma_1=\gamma_2=0.1$. This will make the expected sample size under $H_3$   smaller in comparison with $H_1$ and $H_2$.

The results of the evaluations are presented in Table 2. The efficiency is defined as the ratio between the minimum ESS calculated for Bayesian and for DBC-test, and respectively beween the Bayesian test and the MSPRT. The efficiency of the DBC-test is still very high, while the efficiency of the MSPRT is notably lower.

\begin{table}[!t]
\begin{tabular}{l|rrrrrrrr}
$\alpha$&0.1 &0.05 &0.025 &0.01 &0.005 &0.002 &0.001 &0.0005\\
\hline
Bayes&		
121.79&168.73&211.73&264.46& 302.10&350.40&386.22&421.68\\ DBC&122.63&169.58&212.46&264.99&302.69&350.96&386.81&422.18
\\
\hline
 Efficiency (\%)&99.32&99.50&99.65&99.80&99.81&99.84 &99.85&99.88
 \end{tabular}
\caption{Efficiency data for $\theta_1=0.3$, $\theta_2=0.4$, $\theta_3=0.5$, $\gamma_1=\gamma_2=\gamma_3=1/3$}
\end{table}

\begin{table}[!t]
\begin{tabular}{l|llll}
$\alpha$&0.1&0.05&0.01&0.001\\
\hline
 Bayes
&22.93&33.35&54.42&81.37
\\
DBC&23.49 (97.62\%)&33.59 (99.30\%)&54.49 (99.87\%)&81.48 (99.87\%)\\
MSPRT&27.04 (84.80\%)&38.02 (87.72\%) &58.82 (92.52\%)& 85.74 (94.90\%)\\

 \end{tabular}
\caption{Efficiency data for $\theta_1=0.1$, $\theta_2=0.3$, $\theta_3=0.5$, $\gamma_1=\gamma_2=0.1$, $\gamma_3=0.8$}
\end{table} \vspace{4pt}

\vspace{4pt}

{\flushleft 
{\em Example II} \vspace{2pt}

Let us consider a case of group sequential testing for normal observations. Let $Y_i$, $i=1,2,\dots, M$ be samples from a normal $N(m,1)$ distribution being extracted in a group sequential setting (starting from one sample and sequentially taking up to a maximum number of $M$ samples. Let us suppose each sample consists of the same number $n$ of observations. This is a typical model for group sequential hypothesis testing  in clinical trials \citep[see, for example][]{eales}. The hypotheses of interest are $H_1: m=-\delta$ vs. $H_2: m=\delta$, $\delta=0.1$. The  problem is to find a sequential plan with error probabilities of the first and second type equal to $\alpha=\beta$ minimizing a weighted expected number of observations (accordinig to criteria $F_1,F_2,F_3,F_4,F_5$ in \cite{eales}).  We choose $F_4$ as a criterion of minimization for our numerical example, just because no evaluation data for this is shown in \cite{eales}. In terms of our setting in Section \ref{Lagr}, we have $\theta_1=-\delta$, $\theta_2=\delta$, $k=2$, $K=9$, $\vartheta_i=\delta(i-5)/2$, $i=1,\dots,9$, and $\gamma_i=0.1$ for $i=1,2,3, 4$ and for $i=6,7,8,9$, and $\gamma_5=0.2$ \citep[see the Bayesian setup for 
$F_4$ in ][p.15]{eales}.
We take a group sequential setting with $M=10$ groups and $n=40$ observations per group.
We can use for the evaluation of the optimal  group sequential test, with a slight adaptation, the program in \cite{continuous}, applying the Nelder-Mead optimization (over $\lambda_1=\lambda_2$) in order to fit the error probabilities to $\alpha=\beta=0.05$. The weighted ESS found for the optimal test  is 149.0.

For DBC version we do not need the program  part  of test evaluation, and only use the program for computation of the error probabilities and expected sample size.  Running the Nelder-Mead optimization over $\lambda_1=\lambda_2$ again,  we find a DBC-test providing the best fit to $\alpha=\beta=0.05$. The weighted ESS found for this is 149.75.  So the efficiency, with respect to the optimal test, reached by the  DBC-test   is approx. 99.5.

Taking into account that the fixed sample size (FSS) for $\alpha=\beta=0.05$ and $\delta=0.1$ is 270.55, and the maximum sample size is 400, we are situated at $t=400/270.55=1.48$ in the terms of Table 1 in \cite{eales} while the calculated value of $F_4/FSS=149/270.55=0.55$, i.e. practically as low as  the respective ratio for criterion $F_2$ (which is equal to 0.544 for $t=1.5$, according to Table 1 in \cite{eales}).

It seems that there is a good reason not to include the data for $F_4$ in Table 1, being those as close as this to $F_2$.

{\flushleft \em Example III} \vspace{2pt}

Now, let us apply our results to the case of testing a simple hypothesis vs. a two-sided alternative. 

\cite{Billard} proposed a  computer-oriented method of construction of  ``partial sequential tests'' for testing a simple hypothesis vs. a two-sided alternative in the Bernoulli model. In this example we want to show how  our multi-hypothesis tests can be used for construction of two-sided tests in the case of two hypotheses and to compare their efficiency with respect to the Billard's partial sequential test.

The hypotheses of interest, about  the success probability $\theta$ in a Bernoulli scheme, are $H_0^\prime:\theta=0.5$ vs. $H_1^\prime:\theta=0.2$ or $\theta=0.8$.
The desired error probabilities $\alpha$ and $\beta$ are  equal to $0.05$. 

Let us consider an auxiliary  problem of testing $H_1:\theta=\theta_1=0.2$, $H_2:\theta=\theta_2=0.5$ and  $H_3:\theta=\theta_3=0.8$.
Let us suppose there is a (non-randomized) test $\langle\psi,\phi\rangle$ with 
\begin{eqnarray}
\alpha_{21}
(\psi,\phi)+\alpha_{23}
(\psi,\phi)\leq\alpha,\nonumber\\
\alpha_{12}
(\psi,\phi)+\alpha_{13}
(\psi,\phi)\leq\beta,\label{6-2}\\
\alpha_{32}
(\psi,\phi)+\alpha_{31}
(\psi,\phi)\leq\beta.\nonumber
\end{eqnarray}

Then let us define a test $\langle\psi^\prime,\phi^\prime\rangle$ for testing $H_0^\prime$ vs. $H_1^\prime$
in the following way: let  $\psi_n^\prime\equiv\psi_n $, and $\phi_n^\prime=1$ (rejecting $H_0^\prime$) when either $\phi_n^1=1$ or $\phi_n^3=1$, being $\phi_n^\prime=0$  otherwise (accepting $H_0^\prime$), for all natural $n$.
Then $\alpha(\psi^\prime,\phi^\prime)\leq \alpha $ and $\beta(\psi^\prime,\phi^\prime)\leq \beta$.

Thus, to find a two-sided test for testing $H_0^\prime$ vs. $H_1^\prime$ it suffices to find a test
for three hypotheses, $H_1,\,H_2$ and $H_3$, minimizing the Lagrangian function corresponding to \eqref{6-2}:
\begin{eqnarray}\label{6-1}
& &\lambda_{21}(\alpha_{21}
(\psi,\phi)+\alpha_{23}
(\psi,\phi))+
\lambda_{12}(\alpha_{12}
(\psi,\phi)
+\alpha_{13}
(\psi,\phi))\nonumber\\
&&\quad+\lambda_{32}(\alpha_{32}
(\psi,\phi)+\alpha_{31}
(\psi,\phi))+C_{\gamma,\theta}(\psi),
 \end{eqnarray}
with an appropriate choice of $\lambda_{21},\lambda_{12},\lambda_{32}$ and any convenient $\gamma$.

We used the algorithms developed in \cite{novikovmult}
to 
numerically find a test minimizing \eqref{6-1} with $\alpha$ and $\beta$ as close as possible to 0.05, using the Nelder-Mead optimization.   $\gamma_1=\gamma_3=0.25$ and $\gamma_2=0.5$  were used for all numerical evaluations in this example. The results of the evaluations are reported in Table 3 in the  ``Optimal''  column: OC stands for the operating characteristic and ESS for the expected sample size. 

Another evaluation was made, under the same scheme, with our DBC-test \eqref{1-7} -- \eqref{3-1}. The results are reported in Table 3  in the ``DBC'' column. The performance results for both evaluations are  close to each other, with slightly lower ESS values but larger error probabilities in the DBC part.  

For comparison,  the results of Monte Carlo simulations for the ``partial sequential'' test from \cite{Billard}  are also placed in Table 3. The ESS levels for this test notably exceed those of the  two other tests, but this can be  explained, at least in part, by lower levels of error probabilities  this latter test   demonstrated.
It is remarkable that ``theoretical''  values for the OC of Billard's test are quite close to those  our ``optimal'' test really has, while our `optimal'' test shows even better NSS levels than the Billard's 
optimal test theoretically designed for (see Table 1 in \cite{Billard}).

The overall conclusion is that the DBC version  performs nearly as well as the optimal test does.

\vspace{4pt}
\begin{table}[!t]
\begin{tabular}{|l|ll|ll|ll|}
\hline
 &\multicolumn{2}{c|}{Optimal}&\multicolumn{2}{c|}{DBC}&\multicolumn{2}{c|}{Partial sequential}\\
 \hline
$\theta$& OC& ESS& OC& ESS& OC& ESS\\
\hline

 0.50& 0.950& 18.44& 0.944& 17.64& 0.974& 20.12\\
 0.55& 0.921& 19.42& 0.914& 18.41& 0.943& 21.26\\
 0.60& 0.816& 21.93&  0.813& 20.39& 0.847& 25.21\\
 0.65& 0.615& 24.36& 0.622& 22.34& 0.640& 29.55\\
 0.70& 0.361& 24.31& 0.378& 22.42& 0.355& 29.34\\
 0.75& 0.157& 21.33& 0.172& 20.06& 0.168& 25.54\\
 0.80& 0.050& 17.36& 0.056& 16.57& 0.038& 20.54\\
 0.85& 0.011& 13.90& 0.012& 13.38& 0.010& 16.69\\
 0.90& 0.001& 11.31& 0.001& 10.99& 0.000& 14.27\\

\hline
\end{tabular}
\caption{Performance comparison of two-sided tests (the data for the partial sequential test   are due to Monte Carlo simulations in \cite{Billard})}
\end{table}


{\flushleft \em Example IV}\vspace{2pt}

Now,  let us examine a case when the observations are independent but have a trend in the mean.

We refer to the example seen in 
\cite{liu} (Case 6). It is supposed that the observations $X_1, X_2,\dots, X_n,\dots$ follow
the normal distribution with mean $E_\theta X_n=\theta n$ and variance 1, where $\theta$ is an unknown parameter.

The hypotheses of interest are $H_1: \theta=0$, $H_2:\theta=-0.2$ and $H_3:\theta=0.1$.
In \cite{liu}, various characteristics of the MSPRT with the thresholds $\log(A_{ij})=4.6$ are evaluated. We reproduce here the matrix of calculated error probabilities  
\begin{equation}\label{6-3} 
\left[
\begin{matrix}
\times &3.6\cdot 10^{-3} &4.4\cdot 10^{-3}\\
 6.5\cdot 10^{-4}& \times & 1.1\cdot 10^{-11}\\
 4.0\cdot 10^{-3}&2.2\cdot 10^{-5}& \times
 \end{matrix}\right]
\end{equation}
(each row is calculated  assuming that the respective $H_i$ is true, $i=1,2,3$, and the column  number $j$ contains the probability to accept the respective $H_j$, $j=1,2,3$. We had to re-arrange some items  because the data corresponding to $H_1$ and $H_2$ in the original matrix in \cite{liu} are interchanged for some reason),

The simulations of our test \eqref{1-7} -- \eqref{3-1} were carried out with $\lambda_1=35, \lambda_2=18, \lambda_3=33$, $\gamma_1=\gamma_2=\gamma_3=1/3$ and $10^6$ replications and gave the following  matrix of error probabilities:
$$
\left[
\begin{matrix}
\times &3.4\cdot 10^{-3} & 4.2\cdot 10^{-3} \\
 6.7\cdot 10^{-4} &\times & 0.0\\
  4.0\cdot 10^{-3} &2.6\cdot 10^{-5} & \times
 \end{matrix}\right]
$$
with the correponding standard errors:
$$
\left[
\begin{matrix}
\times &5.8\cdot 10^{-5}&6.5 \cdot 10^{-5}\\
 2.7\cdot 10^{-5} &\times & 0.0\\
 6.3\cdot 10^{-5} &4.0\cdot 10^{-6} & \times
 \end{matrix}\right]
$$

\begin{table}[!t]
\begin{tabular}{|c|cc|cc|cc|cc|cc|}
\hline
 $\theta$&\multicolumn{2}{c|}{-0.2}&\multicolumn{2}{c|}{-0.1}&\multicolumn{2}{c|}{0.0}&\multicolumn{2}{c|}{0.05}&\multicolumn{2}{c|}{0.1}\\
 \hline
 MSPRT&8.91&(.0018)&13.51&(.0036) &14.34&(.0025)&19.69&(.0057)&14.02&(.0028)\\
 DBC&8.94&(.0018)&13.35&(.0033)&14.36&(.0025)&19.75&(.0057)&14.06&(.0028)\\
 \hline
\end{tabular}
\caption{The expected sample size for the MSPRT and the DBC version of the Bayes tests. The numbers in parentheses are the estimated standard errors of the respective means.}
\end{table}
The results of evaluation of the ESS are shown in Table 4.
 Surprisingly, the two distinct suboptimal tests have almost identical sets of performance  characteristics.  \vspace{6pt}

{\flushleft 
{\em Example V} \vspace{2pt}

In this example we want to revisit the case of the numerical solution of the Kiefer-Weiss problem for three hypotheses presented in \cite{novikovmult}, Section 4.3.

The method proposed in  \cite{novikovmult} is based on the following  multi-hypothesis generalization  of the modified Kiefer-Weiss problem. 
We find $\vartheta_i\in (\theta_i,\theta_{i+1})$ , $i=1,2,\dots,k-1$, $\gamma_i\geq 0$, $\sum_{i=1}^{k-1}\gamma_i=1$, and a test $\langle\psi,\phi\rangle$ that minimizes $C_{\gamma, \vartheta}(\psi)+\sum_{i=1}^k\lambda_i\alpha_i(\psi,\phi)$,   such that  $E_{\vartheta_i}\tau_\psi=\sup_{\theta\in (\theta_1,\theta_k)}E_{\theta}\tau_\psi$ for $i=1,2,\dots,k-1$.  
Then for any test with error probabilities not exceeding $\alpha_{i}(\psi,\phi)$, $1\leq i\leq k$, its maximum ESS is greater than or equal to $\sup_{\theta\in (\theta_1,\theta_k)}E_{\theta}\tau_\psi$. This means, $\langle\psi,\phi\rangle$  solves the Kiefer-Weiss problem for $k$ hypotheses.

In this way  in \cite{novikovmult}, a numerical solution to the Kiefer-Weiss problem was found in a particular case of three hypotheses $\theta_1=0.3$, $\theta_2=0.5$ and $\theta_3=0.7$ about the success probability in the Bernoulli model, with $\alpha_1=\alpha_3=0.037$ and $\alpha_2=0.07$. The maximum of the  ESS  is achieved at $\vartheta_1=0.4026 $ and at $\vartheta_2=0.5974$ and is equal to 56.2. To verify these results one can use the program code in \cite{multihypothesisGit} taking $\lambda_1=\lambda_2=\lambda_3=200$ and $\gamma_1=\gamma_2=0.5$.

In the algorithm above, the optimal  test (the one that minimizes the Lagrangian function) plays the key role. The question arises, whether its simplified DBC-version can provide a reasonably good approximation to the Kiefer-Weiss solution. Let us try this out, just substituting the DBC-version for the optimal test in the algorithm above. In the same numerical routine as in \cite {novikovmult} we find a DBC-test  which achieves
its maximum  ESS at two symmetrical points $\vartheta_1=1-\vartheta_2$, providing the best fit to the error probabilities of the Kiefer-Weiss solution (using the Nelder-Mead optimization). 
The best approximation is obtained with $\lambda_1=\lambda_3=6.582$ and $\lambda_2=5.964$ and $\gamma_1=\gamma_2=0.5$ giving $\alpha_1=\alpha_3=0.0376$ and $\alpha_2=0.0706$. It is impossible to obtain a better match to the error probabilities of the optimal test above due to the discrete nature of the probabilities involved. The maximum value of the ESS is 56.01 and is achieved at $\vartheta_1=1-\vartheta_2=0.4026$. This result may seem paradoxical because the value achieved by a suboptimal test is smaller than that of the (optimal) Kiefer-Weiss solution of 56.2 mentioned above, but in fact it is not, because this latter test   does not strictly comply with  the  error probabilities of the former.

Anyway, the simplified variant of the optimal test provides  an excellent approximate solution to the Kiefer-Weiss problem, very much like Lorden's  2-SPRT does in the case of two hypotheses \citep{Lorden76}

It is worth noting that the DBC-test evaluations are  far much faster than the backward recursion  in the optimal test, so the DCB version may be recommended for the approximation to the Kiefer-Weiss solution, especially when the number of hypotheses is large. 

\section{Conclusions}

In this paper, we proposed a   new class of sequential multi-hypothesis tests which are not based on the backward induction nor  on any type of recursion. It is aplicable whenever a computable expression for the joint density functions is available at any stage of the experiment. 
Numerical comparisons are made with the optimal (Bayes) and the MSPRT by Armitage, in a particular case of sampling from a Bernoulli population.
Analogous comparison is made with the optimal \citep[minimizing a weighted expected sample size, see][]{eales} group sequential test for normal mean.
A new construction of two-sided sequential test  for two hypotheses is proposed and a numerical comparison  is carried out with the "partial sequential" test by \cite {Billard}, in the Bernoulli model. 
A Monte Carlo study of the performance of the proposed test  is conducted in a case of independent observations with a linear trend in the mean. A numerical comparison with the MSPRT is made. 
An application to the multi-hypothesis Kiefer-Weiss problem is proposed which generalizes the Lorden's 2-SPRT. A numerical comparison of its  performance  with that of the Kiefer-Weiss solution  is carried out.
In all the cases, a very good performance of  the proposed test is observed.

\section*{Acknowledgements}

The author thanks the Associate Editor and the anonymous Reviewers for  their valuable comments and suggestions on improvement of an earlier version of the paper.

The author gratefully acknowledges a partial support of the National Researchers System (SNI) by CONACyT (Mexico) for this work.

  \bibliographystyle{tfcad}

\end{document}